\documentstyle[aps,multicol,epsfig]{revtex}
\begin{document}
\twocolumn[\hsize\textwidth\columnwidth
\hsize\csname @twocolumnfalse\endcsname
\title{{Indirect search for dark matter: prospects for GLAST}}
\author{S\'ebastien Peirani, Roya Mohayaee and Jos\'e A. de Freitas Pacheco } 
\address{
Observatoire de la C\^ote d'Azur, BP 4229, F-06304, Nice Cedex 4, France\\
email: peirani@obs-nice.fr ; roya@obs-nice.fr ; pacheco@obs-nice.fr} 

\maketitle


\begin{abstract}
Possible indirect detection of neutralino, through its $\gamma$-ray annihilation 
product, by the forthcoming GLAST satellite
from our galactic halo, M31, M87 and the dwarf galaxies Draco and 
Sagittarius is studied. $\gamma$-ray fluxes are evaluated for two representative
energy thresholds, 0.1 GeV and 1.0 GeV, at which the spatial resolution of GLAST
varies considerably.  Apart from dwarfs, which are
described either by a modified Plummer profile or by a tidally-truncated
 King profile, fluxes are compared for
halos with central cusps and cores. It is demonstrated that substructures,
irrespective of their profiles, enhance the $\gamma$-ray emission only marginally.
The expected $\gamma$-ray intensity above 1 GeV at high galactic latitudes 
is consistent with the residual emission derived from EGRET data if the density
profile has a central core and $m_{\chi} \leq$ 50 GeV, whereas for a 
central cusp only a substantial enhancement would explain the observations.
From M31, the flux can be detected above $0.1$ GeV and 1.0 GeV by GLAST only if
$m_{\chi} \leq$ 300 GeV and if the density profile has a central cusp, case in which
a significant boost in the $\gamma$-ray emission is produced by the central black hole. 
For Sagittarius, the flux
above $0.1$ GeV is detectable by GLAST provided the neutralino mass is
below $50$ GeV. From M87 and Draco the fluxes are always below the sensitivity limit
of GLAST.

\vspace*{0.5cm}
\end{abstract}
]
\section{Introduction}
\label{introduction}

The nature of dark matter (DM) remains a major unresolved problem in 
astroparticle physics. Main evidences for dark matter come 
from astrophysical observations.
The rotation curve of a typical spiral galaxy flattens or slowly rises 
beyond the optical radius which includes most of the galaxy's
luminosity \cite{pw2000,sr2001}. The flattening of the rotation curves indicates
the presence of DM at large radii (a few tens of kpc): a result
which is confirmed, up to scales of
$30$ kpc, by the study of close binary systems \cite{pj1988}. Among other 
evidences is the confinement of hot X-ray coronae
around giant ellipticals which requires a massive 
dark halo \cite{fjt1985,fabbiano1989}. 
At the centre of spirals or giant ellipticals, kinematical data 
indicate a large ratio of the
baryon to dark matter density (B/DM) 
\cite{op1993,ovp1999,mbd1990}.
On the contrary, for dwarf galaxies it seems that 
the presence of a substantial amount of DM in the 
very central regions is required to explain the kinematical data \cite{has2001}.
On large scales, the average mass to light ratio (M/L) for galaxy clusters
\cite{cye1997}, the infall of the Local Group towards the Virgo cluster \cite{dp1983}, 
the velocity dispersion of galaxies \cite{bp1992} and the angular power spectrum
of the cosmic microwave background (CMB) radiation
\cite{spergeletal2003} have been long providing evidences for dark matter. Various CMB 
experiments and distances to type Ia supernovae favor
a flat universe of total density parameter $\Omega_T \approx 1$ and the matter density 
parameter $\Omega_m \approx 0.30$ \cite{perlmutteretal1999,riessetal1998}. The total baryonic 
contribution, fixed either by the 
primordial nucleosynthesis of light elements
or by the secondary acoustic peaks observed by CMB experiments such as 
WMAP is $\Omega_b \approx 0.04$ \cite{spergeletal2003}, 
indicating that on large scales gravitational forces are 
mainly due to non-baryonic matter.
In view of all these evidences, the inevitable question arises :
what is dark matter made of?

Most dark matter candidates which arise in the standard model of particle physics 
are excluded by the present observational
constraints. For instance, neutrinos ($\nu$) which decouple relativistically 
from the primordial plasma and have a high relic abundance
are the primary hot dark matter (HDM) candidates. However, in a HDM Universe,  
small-scale structures, of typical galaxy sizes, are erased
by relativistic streaming, hence neutrinos are ruled out 
as DM candidates. Moreover,  WMAP
data set a robust upper limit $\Omega_{\nu}h^2 < 0.0076$
to their density \cite{spergeletal2003}. Supersymmetric (SUSY) extensions
of the standard model lead 
to many more new DM candidates
such as s-neutrinos, axions, gravitinos, photinos and cryptons \cite{eln1990}. 
Presently, the most plausible SUSY dark matter candidate is 
neutralino ($\chi$) which is the lightest supersymmetric particle.
Neutralino is stable and hence is a candidate relic from Big Bang, 
if $R$-parity quantum number, introduced to avoid a too rapid
decay of proton, is conserved as is the case in the  
Minimal Supersymmetric extension of the Standard Model
(MSSM).
Neutralino is an electrically neutral Majorana fermion whose mass, $m_\chi$,
can range from a few GeV to a few hundreds of TeV.
A lower limit of about  $m_{\chi} \sim 30\, {\rm GeV}$ has been set
by the LEP accelerator \cite{abbiendietal2000}, while 
an upper limit of $m_{\chi} \sim {\rm 340}$ TeV is favored 
theoretically to preserve unitarity \cite{gk1990}. 

In the past few years, major experiments have gone underway for the {\it
direct} or {\it indirect} detection of dark matter particles \cite{fmw2001}. Direct-detection 
experiments, such as DAMA, EDELWEISS, CRESST and IGEX are among many others
\cite{loidl2002}, which basically
measure the energy (up to tens of keV) 
deposited at the detector by the elastic 
scattering of dark matter particles from the detector
nuclei. Direct detection experiments also use the annual modulation of the
signal due to the orbital motion of the 
Earth around the Sun in their DM search. Clearly, the event rate, the search
strategy and data analysis strongly depend on the spatial distribution of 
dark matter and its dynamics in the galactic halo, which are not unfortunately
well-understood. 
For instance, it is not established if dark matter halos are
dynamically-relaxed structures or not.
Whether dark matter is
homogeneously distributed with isotropic velocity distribution or whether
there are inhomogeneities such as local streams, {\it e.g.} like that
manifested through the tidal arms of the Sagittarius 
dwarf \cite{newberg2003,majewskietal2003} is not entirely clear.
The leading trail of Sagittarius could be showering matter down upon the Solar 
System, affecting the local halo density and perturbing 
the velocity distribution \cite{fgn2003}.
The presence of very high density 
structures such as caustics, if survived up to 
the present day, could 
also have dramatic effects on direct searches \cite{stw1997}. 
Moreover, dark halos are generally not at rest and have considerable 
angular momentum \cite{pmp2003}, whose 
vector direction is probably not the same as those of 
the present spin axes of (baryonic) disks \cite{bachw2002}. If this was the case for the
Galaxy, then significant variations in the modulation of the signal would be
expected. All of these are just a few of many uncertainties about properties
of DM in the halos which overshadow the search experiments.

The indirect detection experiments search for    
products of self-annihilation of neutralinos such as energetic leptons, hadrons
and also particles which would emerge in the follow-up hadronization and fragmentation
processes. In addition to 
$\gamma$-ray lines generated through the annihilation channels 
$\chi\bar\chi\rightarrow\gamma\gamma$  and 
$\chi\bar\chi\rightarrow
Z^0\gamma$, the annihilation of two neutralinos also produces a 
$\gamma$-ray continuum as a consequence of the neutral pion decay. 
Besides these energetic $\gamma$s, neutrinos are also
produced either in quark jets ($b\bar b$ interactions) or in the decay
of $\tau$ leptons and gauge bosons. Neutrinos produced in former
process are less energetic than those produced in the latter.   
Neutralinos can be decelerated by scattering off nuclei and then accumulating at 
the centre of the Earth and or at the centre of the Sun (or inside any other 
gravitational potential well), thus increasing the annihilation rate.
However, so far different experiments designed to detect decay products like
high-energy neutrinos, have only managed to set upper limits 
on fluxes coming from the Earth centre or 
from the Sun \cite{ahrensetal2003}. 

The $\gamma$-ray emission on the other hand is expected, for instance, from
the galactic centre and other massive dark matter halos.
Since the annihilation rate depends 
on the square of the local density, the distribution of DM in the halo is of crucial 
importance. The degree of clumpiness, the density profile of a dark matter
halo and in particular whether it has a central cusp or not,  
as well as the presence or absence of a central
supermassive black hole (SMBH) could influence the annihilation rate. Other events such as 
the infall of small satellites, which are not totally
disrupted by tidal forces can locally enhance the $\gamma$-ray emission \cite{to2002}.

The prediction of $\gamma$-ray fluxes requires
two separate inputs: that coming from particle physics for issues such as 
the interaction cross section and the number of photons per annihilation,  
and the input from astrophysics for problems such as the 
spatial distribution of dark matter in potential sources.

Most studies on indirect DM detection explore
various decay channels of neutralino annihilation 
in the huge parameter space of MSSM, consisting of 91 real parameters and 74 phases
(most of them can be absorbed by field redefinitions \cite{hep0312378}). In the
effective MSSM model, where the low-energy parameters are
constrained by accelerator data, 
seven free parameters still remain (the higgsino mass parameter $\mu$,  the gaugino mass
parameter $M_2$, the ratio of the Higgs vacuum expectation values tan$\beta$, the
mass of the CP-odd Higgs boson $m_A$, the scalar mass parameter $m_0$ and the
trilinear soft SUSY-breaking parameters $A_b$ and $A_t$). The computation 
of SUSY particle spectrum is therefore a difficult and sometimes uncertain procedure,
which is often
done using publicly available numerical codes like 
SUSPECT able to explore the full 
parameter space \cite{dkm2002}.
In this work, we take a different approach and assume that the
number of photons produced per annihilation, $Q_{\gamma}$, can be evaluated under the
assumption that the $\chi\bar \chi$ annihilation process is similar to a QCD
jet. In this picture, the evaluation of $Q_{\gamma}$ is rather simple since 
{\it it depends only on one scale parameter, the neutralino mass}.

The astrophysical aspect of the neutralino 
flux calculation, which concerns the dark matter distribution, remains rather uncertain. Numerical
simulations suggest that the density in the central regions of dark halos varies as $\rho \propto
1/r^{\beta}$, where $\beta$ takes on values such as $\beta=1$ \cite{nfw1996} or $\beta
= 3/2$ \cite{mooreetal1999}. However, as we mentioned earlier, 
a cusp profile is not always supported by the
observational data such as
the rotation curves of bright galaxies \cite{pw2000}. Moreover, in order
to avoid the inevitable divergence of the 
$\gamma$-ray emission rate calculated from these cusp profiles, various assumptions
about the central density or cutoff radius are always 
made (see for example \cite{to2002,cm2000,swsty2003}). As a consequence, 
the value of the square of the density integrated along the line of sight 
in the direction of the galactic centre performed by different
authors varies by as much as four
orders of magnitude (see Table 1). 
The density profile and the M/L ratio also depends on the type of
the galaxy, {\it e.g.}, as compared to normal spirals, dwarfs seem to have a
much higher M/L ratio and do not seem to have a central cusp.
Thus, as far as the selection of a suitable source is concerned, many factors
such as the possible density profile of the source, the level of background 
contamination, the distance to the source and the
M/L ratio should be taken into account. 
Different studies have considered the galactic centre direction as a privileged source
due to its high column density. However, the $\gamma$-ray emission from our halo is  
highly contaminated by the local background, mostly produced 
by cosmic ray interactions with the interstellar environment. Background
contamination is less significant for sources outside the galactic plane and
extragalactic sources which 
subtend a small solid angle at the detector.
Because of their proximity, the galactic centre and the
nearby dwarfs, Draco and Sagittarius, offer unique sites for dark matter
detection. Due to their considerably large dark matter content, M31 and
in particular M87, are also
potential $\gamma$-ray sources. These are the sources that we shall consider in
this article.
We have calculated $\gamma$-ray fluxes from the Milky Way, M31 and M87 using two different density profiles: 
the first one is a very steep but non-diverging central profile based on very
recent numerical simulations
\cite{navarroetal2003}, while the second is a 
core-profile (Plummer profile) chosen because the energy distribution, the density
and the potential are known in a closed form. The Plummer profile also leads to
a finite mass and the density does not diverge at the centre.
For Draco, we have used a density profile resulting from a dynamical model whereas
a King profile was adopted for Sagittarius.
In addition to these computations, we have simulated
halos with clumps in order to verify how substructures affect the 
predicted $\gamma$-ray fluxes and have found that
clumpiness enhances the predicted emission only  marginally, in disagreement
with some previous claims \cite{cm2000,begu1999} and in agreement with 
others \cite{swsty2003}.

Unlike most previous studies which estimate fluxes mainly for atmospheric
Cherenkov telescopes (ACTs) such as CELESTE and VERITAS,
in this work, we consider the possibility of detection by the future Gamma-ray Large-Area Space 
Telescope (GLAST), which has various obvious advantages over ACTs, the more
important of which are as follows: i) lower energy threshold, allowing
to probe neutralino masses above $10$ GeV; ii) the background is mainly due to
the diffuse extragalactic emission and iii) the spatial resolution varies with the
threshold energy, allowing to probe the halo density profile.

Our results suggest that GLAST
may detect M31 at energies above $0.1$ GeV and $1.0$ GeV if the neutralino mass is 
less than  $300$ GeV and only if the profile is cusped. Core profiles require
neutralino masses lower then upper limits established by accelerator data. 
Thus GLAST can put constraints not only on the neutralino mass but also on
the density profile, combining the different spatial resolution at different energies.
According to our computations, in spite of having a very massive halo,
emission from M87 cannot be detected either above $0.1$ GeV or $1.0$ GeV. However,
for M87, there are enough evidences for the presence of a central 
SMBH \cite{harmsetal1994,maccheto1997} which can 
form a central dark matter "spike" and which would consequently 
boost the $\gamma$-ray flux by a factor of about 
$200$, producing a detectable signal.

For the dwarf spheroids Draco and Sagittarius, we evaluate the
flux and demonstrate that $\gamma$-ray emission from Draco 
is below the sensibility limit of GLAST,
whereas Sagittarius can be detected above $0.1$
GeV energy threshold, if the neutralino mass is $\leq 50$ GeV. 

This article is organized as follows. In Section \ref{gammarayemission}, we discuss the
$\gamma$-ray emission and the flux equation. The annihilation rate and the  number of
photons produced per annihilation event are 
discussed in Section \ref{SUSY}. In Section
\ref{ASTRO} we discuss the astrophysical parameters required for the flux
calculation. In Section \ref{numericalresults}, we present our numerical
results for emission from the galactic halo , M31, M87, Draco 
and Sagittarius. In Section \ref{clumps}, we demonstrate how 
clumpiness of halos would effect the $\gamma$-ray flux. In Section
\ref{blackhole} we study flux enhancement effects due to the central
SMBH in M31 and M87 and finally in
Section \ref{conclusion}, we summarize our results and conclude our work.

\section{The $\gamma$-ray flux}
\label{gammarayemission}

Let us assume that dark matter halos are constituted by neutralinos: the lightest
supersymmetric particle in many SUSY models (for a review of SUSY dark matter
see \cite{jkg1996}). Neutralino is a linear superposition,
\begin{equation}
\chi=a_{\tilde B}\tilde B+a_{\tilde W}\tilde W+a_{\tilde H_1}\tilde
H_1+a_{\tilde H_2} \tilde H_2,
\end{equation}   
of four mass-eigenstates
where $\tilde B$ and $\tilde W$ are gauginos (the superpartners of electroweak gauge
bosons) and $H_{1, 2}$ are higgsinos. If the sum $a_{\tilde B}^2 + a_{\tilde W}^2 >$ 0.9, the
neutralino is {\it gaugino} type, while it is {\it higgsino} type if that sum is less than $0.1$.  

We further assume that in the early universe,
neutralinos were in thermal equilibrium with the primordial plasma and
that their present density was fixed at the freeze out, {\it i.e.}, at the
temperature $T_*$ when they decoupled from the other particle species. At
freeze out, the annihilation rate is comparable to
the expansion rate of the Universe and moreover neutralinos are
non-relativistic ($kT_* < m_{\chi}c^2$).
After decoupling, neutralinos can annihilate via the following channels
\begin{equation}
\chi\bar \chi \rightarrow l\bar l, q
\bar q, W^+W^- , Z^0Z^0, H^0H^0, Z^0H^0 ,
W^{\pm}H^{\mp}
\end{equation} 
into leptons and anti-leptons ($l\bar l$), quarks and anti-quarks ($q\bar q$),
charged  ($W^{\pm}$ and $H^{\pm}$) and neutral ($H^0$ and $Z^0$) bosons 
respectively.

The decay of neutral pions formed in the hadronization process is the
dominant source of continuum $\gamma$-rays. Besides the continuum
emission, two annihilation channels may produce $\gamma$-ray lines.
The first, $\chi\bar \chi \rightarrow \gamma\gamma$, where the photon
energy is $\sim \, m_{\chi}$ and the second, $\chi\bar \chi \rightarrow
Z^0\gamma$, where the photon energy satisfies $\epsilon_{\gamma} =
m_{\chi} - m_Z^2/(4m_{\chi})$. The latter process is only important
for neutralino masses higher than $\sim 45$ GeV.
 
Since neutralinos are Majorana particles, their density is equal to that of
anti-neutralinos and hence the annihilation rate per unit volume is
\begin{equation}
\varepsilon_{\chi \bar \chi} = 
<\sigma_{\chi \bar \chi}v>\left(\frac{\rho_{\chi}}{m_{\chi}}\right)^2
\end{equation}
where $<\sigma_{\chi \bar \chi}v>$ is the thermally-averaged annihilation
rate, $m_{\chi}$ is the neutralino mass and
$\rho_{\chi}$ is the neutralino matter density. If the 
source is spatially extended, the radiation intensity $I_{\gamma}(r_p)$, measured at a given
projected distance $r_p$ from the centre is 
\begin{equation}
I_{\gamma}(r_p) = \frac{<\sigma_{\chi\bar \chi}v>}{4\pi m_{\chi}^2}
\,\,Q_{\gamma}\,{\cal I}
\label{igamma}
\end{equation}
where
\begin{equation}
{\cal I}=\int \rho_{\chi}^2\left(\sqrt{s^2 + r_p^2}\right) ds
\label{integral}
\end{equation}
is the {\it reduced intensity}, $Q_{\gamma}$ is the number of photons above a certain energy
threshold produced per annihilation and the integral is carried 
along the line of sight. The flux $f_{\gamma}$
within a solid angle $\Delta\Omega$ subtended by the detector 
($d\Omega = 2\pi r_pdr_p/D^2$ for a small angular distance) is
\begin{equation}
f_{\gamma} = \int I_{\gamma}d\Omega = \frac{2\pi}{D^2}\int I_{\gamma}(r_p)P(r_p)r_pdr_p
\label{fgamma}
\end{equation}
where $D$ is the distance to the source and $P(r_p)$ is the 
point-spread function (PSF) of the detector. In the case of GLAST, we 
adopt a Gaussian PSF ($P(r_p) = e^{-r_p^2/2\sigma^2}$) whose width depends 
on the energy. The GLAST project specification gives the width at which $68\%$ 
of the signal is included, which yields 
$\sigma(0.1 GeV) \approx 1.05^o$  and  $\sigma(1 GeV) \approx 0.17^o$.

\section{Neutralino annihilation rate : the values of $<\sigma_{\chi\bar
    \chi}v>$ and $Q_{\gamma}$}
\label{SUSY}

In order to compute the $\gamma$-ray flux from (\ref{fgamma}), different 
physical and astrophysical parameters should be calculated. First, the average 
annihilation rate,
 which will be estimated under the assumptions already mentioned, i.e., that
 neutralinos are initially in thermal equilibrium and decouple at $T_*$, when
the annihilation rate equals the expansion rate. The latter is determined by the
Friedman equation relating the Hubble parameter  to the total energy
density. As we shall see, neutralinos decouple at
temperatures in the range $0.4-70$\,\, GeV, depending on their assumed mass. Once neutralinos
decouple, their comoving number remains almost constant, but their concentration
with respect to photons changes, since other particles which decouple later, produce
a "reheating" of the photons due to entropy conservation. In this case, the ratio
between the temperature after and before a given decoupling depends on the
variation of the effective number of degrees of 
freedom as $(g_{eff, bef}/g_{eff, aft})^{1/3}$, where
\begin{equation}
g_{eff}(T) = \frac{1}{2}\sum (g_B  +  \frac{7}{8}g_F)
\end{equation}
and the sum is over all bosons, B, and fermions, F, present in the
primordial plasma.  For temperatures above $0.4$\,\, GeV and below $80$\,\,  GeV one would 
expect that quarks and gluons are present and that the gauge bosons ($W^{\pm}$ and
$Z^0$) have already disappeared. In this case, the main particles contributing to
$g_{eff}$ are photons, gluons, the three leptons and their anti-particles
(the tauon decouples at $T_* \sim 1.8\,\, GeV$) as well as their associated neutrinos and
anti-neutrinos, the six quarks and their anti-quarks.

Under these conditions one obtains (see for instance reference \cite{kt1990})
the {\it balance equations}
\begin{eqnarray}
\Omega_{\chi}h_{65}^2 &=& \frac{6.88\times 10^7}{g_{eff}(T_*)}m_{\chi}(GeV)
y^{3/2}e^{-y}\,,
\label{balance1}
\\
\langle\sigma_{\chi \bar \chi}v\rangle &=& 
\frac{8.82\times 10^{-36}\sqrt{g_{eff}(T_*)}}{g_{\chi}
m_{\chi}(GeV)}y ^{-1/2}e^{y}\,\, cm^3s^{-1}\,,
\label{balance2}
\end{eqnarray}
where $y = m_{\chi}c^2/kT_*$ and $h_{65}$ is the Hubble parameter in units of
$65$ kms$^{-1}$Mpc$^{-1}$. From these equations, once the density parameter is fixed,
the decoupling temperature and the annihilation rate per particle can be computed
for a given neutralino mass. Here we adopt the
cosmologically accepted value of $\Omega_{\chi} = 0.26$
and assume that the neutralino mass lies in the 
range 10 GeV$ \leq m_{\chi} \leq$ 2000 GeV. In this case, the
decoupling temperature parameter $y$ varies within the interval 
$22.5 \leq y  \leq 28.2$, corresponding to the temperature range mentioned above.
The mean annihilation rate per particle varies very little, $ (7.7 - 9.5) \times 10^{-27} \, {\rm
  cm}^3{\rm s}^{-1}$  and is comparable to values adopted in
other studies \cite{to2002,salati2002}.

A major difficulty in the calculation of the number of $\gamma$s per
annihilation stems from the unclear details of the hadronization
and gamma-ray emission in the large-dimensional parameter space of SUSY models. 
Here, we use an approximation to the numerically computed  QCD fragmentation functions
taken from  \cite{hsw1987,wmb2002}.
We assume that hadronization and subsequent $\gamma$-ray emission is similar to
the decay of QCD jets and that 
the annihilation of two neutralinos results in two jets, each
of energy $\sim m_{\chi}$.  Using the fragmentation function of jets into photons, neutrinos
and baryons, which is known to a good approximation up to a few TeV
\cite{hsw1987,hill1983}, the number of photons per annihilation and per jet
produced in the energy interval $x, x+dx$ is
\begin{equation}
\frac{dQ_{\gamma}}{dx} = \frac{5}{8}\left(\frac{16}{3} 
-2\sqrt{x}-\frac{4}{\sqrt{x}}+\frac{2}{3}x^{-3/2}\right)
\end{equation} 
where the dimensionless energy is defined as $x = \epsilon_{\gamma}/(m_{\chi})$,  with
$\epsilon_{\gamma}$ being the photon energy. Therefore, the total number of photons per
annihilation event above the threshold energy $\epsilon_{\gamma, {\rm th}}$ is simply given by 
\begin{equation}
Q_{\gamma} = 2\int_{x_{\rm th}}^1 \frac{dQ_{\gamma}}{dx}dx\,.
\end{equation}
where the factor two takes into account the formation of two jets
rather than one, due to momentum conservation.

\begin{figure}
\centerline{
\epsfxsize=0.5\textwidth
       {\epsfbox{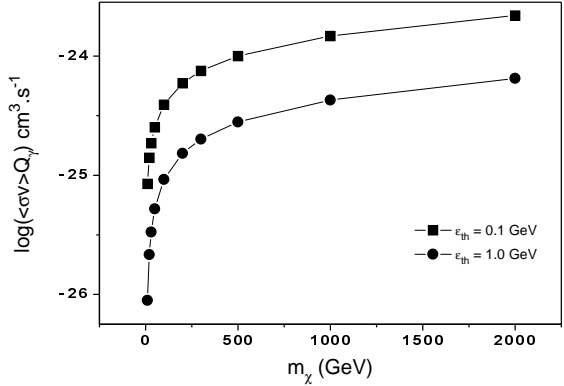}}
}
\vspace*{0.1cm}
\caption
{
\noindent
The variation with the neutralino mass of the photon production rate per annihilation 
above the energy thresholds, 
$\epsilon_{\gamma, {\rm th}}$,  $0.1$ GeV and
$1.0$ GeV.}
\label{jets}
\end{figure}

In Fig.\ \ref{jets}, we show the photon production 
rate $<\sigma_{\chi\bar \chi}v>Q_{\gamma}$ 
in ${\rm cm}^3{\rm s}^{-1}$ for two threshold energies ($0.1$ and $1.0$ GeV) as a 
function of the 
neutralino mass. The rates, in spite of our simplifying assumptions, are in
good agreement with estimates derived from SUSY numerical 
codes \cite{to2002,salati2002}. This is not  
surprising since the parameters required in SUSY codes are chosen to match the expected
relic density. We have taken a reversed approach, {\it i.e.}, we fix the relic density
first and then compute the annihilation rate using the
balance equations (\ref{balance1}) and (\ref{balance2}).
As a consequence, both procedures should give comparable results, but our method has
the advantage of using only one free parameter, the neutralino mass. 

\vspace*{-0.4cm}

\section{Astrophysical parameters: the values of ${\cal I}$}
\label{ASTRO}

In the previous section, we have evaluated the annihilation rate $<\sigma_{\chi \bar \chi}v>$
and the total number of photons per
annihilation event above a given energy threshold, $Q_\gamma$. In this section, we
evaluate the reduced intensity ${\cal I}$, given by (\ref{integral}) whose value is needed
for the evaluation of the $\gamma$-ray flux (\ref{fgamma}).
The integral in (\ref{integral}) is often evaluated assuming a density
profile of the general form
\begin{equation}
 \rho(r)  = \frac{\rho_*}{(r/r^*)^{\beta}(1+ r/r_*)^{\gamma}}
\label{profile}
\end{equation}
where $\rho_*$ and $r_*$ are the characteristic density and radius.
If $\beta =1$ and $\gamma = 2$, one obtains the so-called 
Navarro-Frenk and White (NFW) profile \cite{nfw1996}, while if
$\beta = \gamma = 1.5$ one obtains the so-called Moore profile \cite{mooreetal1999}.
Clearly, when integrated along the line of sight passing through the centre, as required by 
expression (\ref{integral}), both of these two profiles diverge.
Various tricks are used to overcome this difficulty. One can impose a
finite central density  such as $\rho_{\chi} = m_{\chi}H_0/<\sigma_{\chi\bar \chi}v>$,
since denser regions will be depleted by high neutralino annihilation rates \cite{cm2000}.
High resolution simulations are also used to study the behaviour of the density profile 
in the very central regions of halos. However, even the {\it highest} resolution 
simulations are presently unable to resolve the very inner structure of halos 
and different extrapolations are needed to
obtain the central density. In order to illustrate this point, Table 1 
compares different values of
density-squared along the direction of the galactic centre given by the
expression (\ref{integral}) estimated by 
several authors. In fact, these reduced intensities correspond to values {\it averaged}
over a given solid angle representative of the spatial resolution of the detector.

\vspace*{1.0cm}
\noindent
{\small Table 1. Reduced intensity in the direction of the galactic centre}
\begin{flushleft}
\begin{eqnarray}
{\rm Profile}  & \qquad\quad \int \rho^2ds\,\, ({\rm GeV}^2{\rm cm}^{-5}) \qquad   {\rm
  Reference} \nonumber\\
\hline\nonumber\\
{\rm Moore}   & \qquad\qquad  3.3\times 10^{26}  \qquad\qquad\qquad\,\, \cite{cm2000} \nonumber\\
{\rm NFW}  & \qquad\qquad  2.8\times 10^{25}  \qquad\qquad\qquad\,\,  \cite{efs2003} \nonumber\\
{\rm  core}  & \qquad\qquad 3.0\times 10^{22} \qquad\qquad\qquad\,\,  \cite{efs2003} \nonumber\\
{\rm cusp}  & \qquad\qquad  2.4\times 10^{22}  \qquad\qquad \qquad\,\, \cite{ellisetal2001} \nonumber\\
{\rm NFW} &  \qquad\qquad  5.2\times 10^{25}   \qquad \qquad \qquad\,\, \cite{swsty2003}\nonumber\\
{\rm SWTS}  & \qquad\qquad  1.8\times 10^{24}  \qquad\qquad \qquad\,\, \cite{swsty2003} \nonumber
\end{eqnarray}
\end{flushleft}

\noindent
Simple inspection of Table 1 shows that the 
integral (\ref{integral}) may vary by four orders
of magnitude according to the assumed profile. Consequently, severe
uncertainties overshadow the value of the predicted $\gamma$-ray flux.

Rather recently, the inner structure of dark halos in $\Lambda$CDM cosmology was
revisited \cite{navarroetal2003} and the new  ''$\alpha$-profile''
\begin{equation}
\rho(r) = \rho_* {\rm exp}\left(-{2\over \alpha}
\left[\left({r\over r_*}\right)^{\alpha} - 1\right]\right)
\label{alpha}
\end{equation}
fitting the the numerical results 
was proposed, which has several advantages over 
(\ref{profile}). Firstly, the total mass and the central density are finite
which solve the aforementioned divergence problem in the 
evaluation of the central intensity. Secondly, the logarithmic slope
decreases inwards more gradually than NFW or Moore profiles. 
According to simulations, the parameter $\alpha$ is 
restricted to be in the range $0.1-0.2$ and here we adopt for M31 and the Milky Way
the suggested value of $\alpha = 0.17$  \cite{navarroetal2003}.
The two remaining parameters, $\rho_*$ and $r_*$, can 
be estimated if the total halo mass and one other
physical quantity such as the maximum circular velocity are known.  
However, the rotation curve within the effective
radius {\it is not dominated by dark matter} \cite{pmp2003} and so the maximum rotation velocity is not
a robust quantity. The halo of M31, as we shall see, is probably similar to our 
own halo which implies that the dark matter energy density at a distance of about $8-9$ kpc
from the centre should be comparable to that observed in the solar vicinity, {\it i.e.}
$0.4$ GeVcm$^{-3}$. We use this value together 
with the total mass of M31 to fix the parameters $\rho_*$ and $r_*$. 

Present observational data constraining the mass of the halo of M31 are 
rather poor. For instance, the
rotation curve seems to be well-explained with just the baryonic matter in the 
bulge and the disk, whose mass distributions are traced by the blue 
light \cite{braun1991}. This does not exclude the existence of a dark matter halo
whose contribution to the gravitational force, and hence its effect on the
rotation curve, would be more relevant only at distances 
larger than few tens of kpc.  Direct estimates of the halo mass of M31 are based 
on the peculiar motion of the Local Group (LG) and on the kinematics of its satellites.
A new solution for the solar motion which includes the $32$ probable
members of the Local Group and assumes virial equilibrium, leads to a total
mass for the LG of $(2.3\pm 0.6)\times 10^{12} M_\odot$ \cite{cb1999}.
Using the projected
mass method (\cite{bt1981,htb1985}), the contribution of the M31 
subgroup (including only 7 satellites) to the total mass of the LG amounts to 
$(13.3\pm 1.8)\times 10^{11}\, M_\odot$ \cite{cb1999}. A similar result
($12.3^{+18}_{-6.0}\times 10^{11}\, M_{\odot}$) was
obtained \cite{ew2000}, including $10$ satellites, $17$ distant globular clusters
and $9$ halo planetary nebulae, while a firm upper bound of $24\times 10^{11}\ M_{\odot}$
was derived from a similar analysis including $15$ satellites and new radial velocity data
\cite{ewggv2000}. Using the kinematical data on these $15$ satellites and the projected
mass estimator \cite{htb1985}, we have obtained a mass of $(1.5\pm 0.5)\times
10^{12}\, M_{\odot}$ for M31. Other indirect estimates based on disk formation and cusp
dark matter halo expected for $\Lambda$CDM cosmology predict a mass of
$1.6 \times 10^{12}\, M_{\odot}$ \cite{kzs2002}. We adopt the value
$1.5 \times 10^{12}\, M_{\odot}$, which is compatible with all of 
these different estimates. 

Subsequently, using the mass of M31 and also the local density of $0.4$ GeVcm$^{-3}$
which we argued for earlier in this section, we obtain 
the two parameters defining the ''$\alpha$-profile''
(\ref{alpha}): $\rho_* $ = $4.5\times 10^{-25}\, {\rm gcm}^{-3}$ and
$r_* = 11.6$ kpc. We emphasize that these  parameters give, in terms
of the total potential energy $\mid W\mid$, a gravitational
radius of
$r_g = GM^2/\mid W \mid \approx 250$ kpc, which is a typical value that we
have found in our numerical simulations for halos of similar masses \cite{pmp2003}.

Since the high B/DM ratios observed in the central regions of galaxies
is contradictory with a cusp in the DM density profile, we have also considered a 
profile with a central core to represent the dark matter
distribution inside halos. We assumed a Plummer density profile, 
\begin{equation}
\rho(r) = \frac{\rho_0}{[1+(1/3)(r/r_0)^2]^{5/2}},
\label{plummer}
\end{equation}
and since the central density and total mass are finite, the distribution function
depends only on the total energy ( $f(E) = K\mid E \mid^{7/2}$) and the gravitational
potential can also be obtained in a closed form. 
Using the two constraints mentioned earlier in this section, we obtain the 
two parameters defining the profile (\ref{plummer}) in the case of M31: 
$\rho_0$ = $2.8\times 10^{-24}\, {\rm gcm}^{-3}$ and $r_0 = 12.2$ kpc. 

For M87, the properties of the gaseous X-ray corona were determined in reference
\cite{tsai1993}. From the density and temperature profiles of the hot gas, the total 
mass inside a given radius $r$ can be computed under the assumption of
hydrostatic equilibrium. After correction for the contribution of baryons (stars and gas)
to the gravitational forces, the
dark matter distribution can be evaluated \cite{tsai1993}. Remarkably, such a 
distribution can be well-represented by an  ''$\alpha$-profile'' with the
parameters $\alpha= 0.20, \rho_* = 1.07\times 10^{-25}\, {\rm gcm}^{-3}$ and
$r_* = 79.9$ kpc. Integration of this profile gives a mass
of $9.7 \times 10^{12}\, M_{\odot}$ inside a radius of $100$ kpc, in agreement with other
estimates \cite{tsai1993}. However, we cannot exclude that at distances less
than a few kpc, such a profile would be inconsistent with  observations.

\vspace*{1.0cm}
\noindent
{\small Table 2. Values of the  parameters used in the density profiles.} 
\begin{eqnarray}
\hline\nonumber\\
&\hspace{-3.5 true cm}{\underline{\rm M31}}\quad  &\nonumber\\
&\hspace{-0. true cm}{\rm cusp} : \qquad \alpha = 0.17\qquad  &  \rho_*=4.5\times 10^{-25}\,\,{\rm {g/ cm^3}} \nonumber\\
       &                            &    r_*=11.6 \,\, {\rm kpc}                      \nonumber\\
       &                            &                                                 \nonumber\\
&\hspace{-0.7 true cm}{\rm core} :\qquad\qquad\qquad  & \rho_0=2.8\times 10^{-24}\,\,{\rm {g/ cm^3}}  \nonumber\\
       &                & r_0 = 12.2\,\, {\rm kpc}                          \nonumber\\
\hline\nonumber\\
&\hspace{-3.5 true cm}\underline{\rm M87}\quad   &\nonumber\\
&\hspace{-0. true cm}{\rm cusp} : \qquad \alpha = 0.20\qquad & \rho_*=1.07\times 10^{-25}\,\,{\rm {g/ cm^3}} \nonumber\\
       &                &      r_*=79.9 \,\, {\rm kpc}                      \nonumber\\
\hline\nonumber\\
&\hspace{-3.4 true cm}\underline{\rm Draco}\quad  &\nonumber\\
& \hspace{-0.5 true cm}{\rm Modified\,\,Plummer\,\, profile} :   &                   \nonumber\\
                    &     &\beta = -0.34\,\,        \nonumber\\  
                    &     & \gamma = -1.03\,\,       \nonumber\\
                    &    & \rho_0 =3.8\times 10^{-23}\,\,{\rm {g/ cm^3}} \nonumber\\
\hline\nonumber\\
&\hspace{-2.8 true cm}\underline{\rm Sagittarius}\quad  &\nonumber\\
& \hspace{-1.7 true cm}{\rm King\,\, profile} :   &                   \nonumber\\
                    &    & r_{\rm t}=2\,\, {\rm kpc} \nonumber\\
                    &    & r_{\rm c}=0.55\,\, {\rm kpc}      \nonumber\\
                    &    & \rho_0=2.2\times 10^{-24}\,\,{\rm {g/ cm^3}} \nonumber\\
\end{eqnarray}

\begin{figure}
\centerline
{
\epsfxsize=0.5\textwidth
{\epsfbox{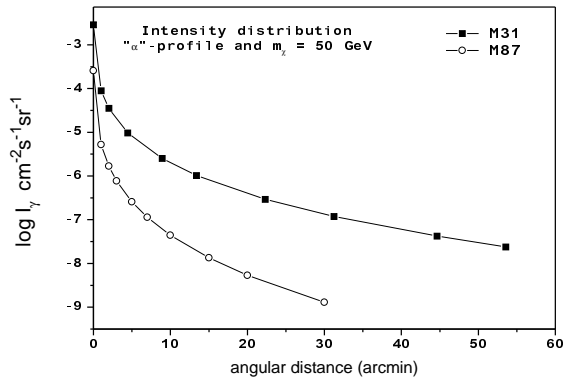}}
}
\vspace{0.3cm}
\caption
{Intensity profiles for M31 and M87 for photon energies above $0.1$ GeV. 
Intensities were computed using the cusp density profile (\ref{alpha}) and $m_{\chi}$ =
$50$ GeV }
\label{intensityprofile}
\end{figure}

These parameters, summarized in Table 2 (for dSph's see details in 
Section \ref{numericalresults}), will be used in our predictions of the
expected $\gamma$-ray flux from neutralino annihilation in M31 and M87.
In Fig. \ref{intensityprofile} we show the expected intensity profile for M31 and M87 as a function of
the angular distance to the centre. These profiles were derived numerically from 
(\ref{igamma}), for an energy threshold of $0.1$ GeV and $m_{\chi} = 50$ GeV. The
comparison uses only density profiles with central cusps, defined by the parameters discussed above.
Note that although M31 has a less massive halo, its central brightness is 
expected to be higher than that of M87. 

\vspace{-0.4cm}

\section{Numerical results}
\label{numericalresults}

\subsection{Flux from the galactic halo}

It is instructive to compare our results with previous estimates of the
reduced intensity (\ref{integral}) in the direction of the galactic centre.
In our work, the integral of the density squared along the line of sight
is $1.6 \times 10^{26}\, {\rm GeV}^2 {\rm cm}^{-5}$ if the $\alpha$-profile
(\ref{alpha}) is adopted. This compares
quite well with the value of \cite{cm2000}, based on extrapolation from high resolution
simulations, but is one order of magnitude higher than those given in a few
of the previous works 
\cite{swsty2003,efs2003}, where NFW profiles were adopted.
Using the core profile, this integral (\ref{integral}) 
is equal to $1.2\times 10^{23}\, {\rm GeV}^2 {\rm cm}^{-5}$, 
considerably smaller than estimates for profiles with central cusps 
as would be expected, but still one order of magnitude higher than
some others which also consider a core profile (see for example
\cite{efs2003} whose estimate of ${\cal I}$ is given in Table 1).

In the energy range $0.1 - 1.0$ GeV, cosmic ray interactions give a substantial contribution
to the $\gamma$-ray emission originated from  the galactic disk. The electron component
produces high energy photons either by inverse Compton scattering or
bremsstrahlung, while
protons produce $\gamma$-photons via the decay of neutral pions 
generated in collisions with interstellar matter. 
In a previous work which uses EGRET data, at high galactic latitudes
a significant residual intensity of $10^{-7} - 10^{-6}$
{\rm photons}\,\,${\rm cm}^{-2}{\rm s}^{-1}{\rm sr}^{-1}$ above $1$ GeV is
claimed even after correction for the expected background of 
cosmic rays and the diffuse extragalactic emission 
\cite{dixonetal1998}. These residual intensities although marginal
have rather interesting implications for the halo density profile. 
We computed the expected intensity, averaged over cells of
$0.5^o \times 0.5^o$ as considered in \cite{dixonetal1998}, in the direction
$b = 90^o$. Table 3 gives the expected intensity as a function of the
neutralino mass for the $\alpha$-profile (with a central cusp) and
the Plummer profile (with a central core).

\vspace*{0.5cm}
\noindent
{\small Table 3. Intensity (in photons $cm^{-2} s^{-1} sr^{-1}$) above 1 GeV in the
direction $b = 90^o$}
\begin{flushleft}
\begin{eqnarray}
&\hspace{-0.41cm}{\rm m_{\chi} (GeV)}  & \qquad \alpha{\rm -profile} \qquad\quad {\rm
  Plummer}\,\,{\rm  profile} \nonumber\\
\hline\nonumber\\
&\hspace{-1cm} 30   &  \qquad  1.1\times 10^{-8}  \qquad\qquad  \,\,1.9\times 10^{-7} \nonumber\\
&\hspace{-1cm} 50   &  \qquad  6.4\times 10^{-9}  \qquad\qquad  \,\,1.1\times 10^{-7} \nonumber\\
&\hspace{-1cm} 100  &  \qquad  2.8\times 10^{-9}  \qquad\qquad  \,\,4.9\times 10^{-8} \nonumber\\
&\hspace{-1cm} 200  &  \qquad  1.1\times 10^{-9}  \qquad\qquad  \,\,2.1\times 10^{-8} \nonumber\\
&\hspace{-1cm} 300  &  \qquad  6.8\times 10^{-10} \qquad\qquad  1.2\times 10^{-8}\nonumber\\
&\hspace{-1cm} 500  &  \qquad  3.4\times 10^{-10} \qquad\qquad  6.1\times 10^{-9} \nonumber\\
\nonumber
\end{eqnarray}
\end{flushleft}
The results summarized  in Table 3 show that at high
latitudes, contrary to what is obtained in the direction of the galactic centre, the
Plummer profile produces an intensity {\it higher} than that derived from the
$\alpha$-profile. This is easily  understood since the latter profile gives a larger
mass concentration near the centre while the former has a shallower mass distribution.
Furthermore, Table 3  shows  that intensities derived
from the $\alpha$-profile are always below the EGRET residual intensity. In this case, 
an important intensity enhancement by
clumps is necessary to explain the EGRET residuals
\cite{begu1999,cm2000}. Intensities
derived from the Plummer profile are comparable to EGRET values if
$m_{\chi} <$ 50 GeV, a limit not inconsistent with the lower bound derived
from LEP. As we shall see in Section \ref{clumps}, the enhancement  of
the $\gamma$-ray emission produced by subhalos is rather small, with
boost not exceeding a factor of $2$. Presently, a firm conclusion
cannot be made since the EGRET residuals are in the sensibility limit of the
instrument. This situation is expected to improve greatly with the forthcoming GLAST.

\subsection{Fluxes from M31 and M87}  

The detection of $\gamma$-rays with energies around 50 GeV from
M31 was considered by the CELESTE collaboration
\cite{falvardetal2002}.  
Their study shows that the annihilation signal
could be detectable by CELESTE
for $ m_{\chi} \geq$ 200 GeV, if the M31 halo is very clumpy and/or if there is
a strong accretion onto a central supermassive black hole \cite{falvardetal2002}.

M87, the giant elliptical galaxy at the centre of the Virgo cluster, is also
a potential source of $\gamma$-rays produced by $\chi\bar \chi$ annihilation 
\cite{bbsts1999}.
If it is simply assumed that the neutralino mass is
$1$ TeV and that the photon production rate above the threshold energy of $50$ GeV is  
$<\sigma_{\chi\bar \chi}v>  Q_{\gamma} = 1.5\times10^{-24}\, {\rm cm}^3{\rm
 s}^{-1}$, the predicted $\gamma$-ray flux from M87 would be below the sensibility of 
the present  ACTs \cite{bbsts1999}. However, detection by the new generation of ACTs like VERITAS 
would be possible if there is an enhancement by
a factor $40$ due to the clumpiness of the halo \cite{bbsts1999}.  
In this work, we consider the possibility of detection by GLAST, which has various obvious
advantages over the atmospheric Cherenkov telescopes, which we have mentioned
in Section \ref{introduction}.

The photon production rate above a given energy is
\begin{equation}
R_{\gamma} = <\sigma_{\chi\bar \chi}v>Q_{\gamma}=
\frac{4\pi f_{\gamma, {\rm min}}}{\int P(r_p)d\Omega\int \rho_{\chi}^2 ds}m^2_{\chi}
\label{pmin}
\end{equation}
where $f_{\gamma,{\rm min}}$ is the minimum flux above a certain photon energy detectable by
GLAST and we have used equations \ref{igamma} and \ref{fgamma}. Values of $f_{\gamma,{\rm min}}$ varies from one  $\gamma$-ray experiment to
another and here, we use the sensibility curves summarized in \cite{hsfp1999}. 
The production rate $R_{\gamma}$ given by the above expression represents the minimum
required value for the production of a detectable signal. In Fig.\
{\ref{m31glast1} we plot the minimum photon production rate
and the photon production rate derived in Section \ref{SUSY} as a function of the neutralino mass, for
an energy threshold of $0.1$ GeV. The required minimum production rates are shown for both
the $\alpha$-profile and the Plummer
profile. Numerical calculations were performed adopting a distance of $770$ kpc for M31. 
\vspace{-0.40cm}
\begin{figure}
\centerline
{
\epsfxsize=0.5\textwidth
{\epsfbox{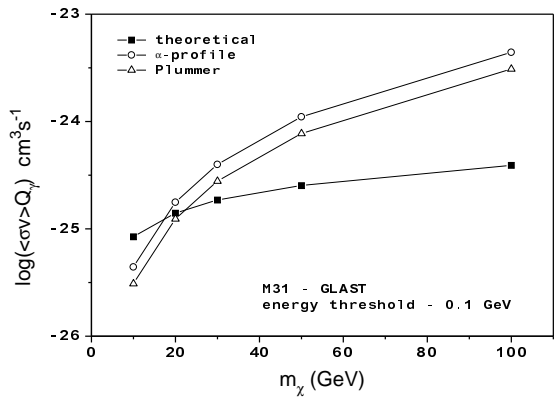}}
}
\vspace{0.2cm}
\caption
{
Minimum photon production rate required for detection and theoretical estimates 
as a function of the neutralino mass and for an energy 
threshold of $0.1$ GeV. Minimum production rates
are shown for the halo of M31 with cusp ($\alpha$-profile )
and core (Plummer) profiles.}
\label{m31glast1}
\end{figure}

Fig.\ \ref{m31glast1} demonstrates that detectable fluxes from M31 can be produced by both
density profiles if   $m_{\chi} \leq 25$ GeV. In fact, the relatively
large acceptance angle of GLAST at $100$ MeV, includes a substantial fraction of the
halo, reducing the expected differences due to different density profiles. The situation
is rather different when the threshold energy is $1.0$ GeV (Fig. \ref{m31glast2}).
In this case, the detector acceptance angle is considerably reduced and
only a central cusp would lead to enough projected mass to produce a
detectable flux if $m_{\chi} \leq 20$ GeV. These mass limits are slightly smaller
than the lower bound derived from LEP \cite{abbiendietal2000}, but masses as low 
as $6$ GeV have been derived from WMAP data in the frame of an effective MSSM model
without gaugino-mass unification at the GUT scale \cite{bdfs2003}. 
In Section \ref{blackhole} we discuss how
these limits are changed by the presence of a central SMBH.

In spite of its extremely massive halo, the predicted fluxes from M87 
are about one order of magnitude below the sensibility limit of GLAST in the
range $0.1 - 1.0$ GeV, due to its large distance ($17$ Mpc) from us and 
in spite of the fact that a substantial
fraction of a far halo would lie within the acceptance angle of the detector. 
As in the case of M31, the central SMBH strengthens considerably the emission,
producing a detectable signal, as we shall shortly demonstrate in Section \ref{blackhole}.

\vspace{-0.4cm}
\begin{figure}
\centerline
{
\epsfxsize=0.5\textwidth
{\epsfbox{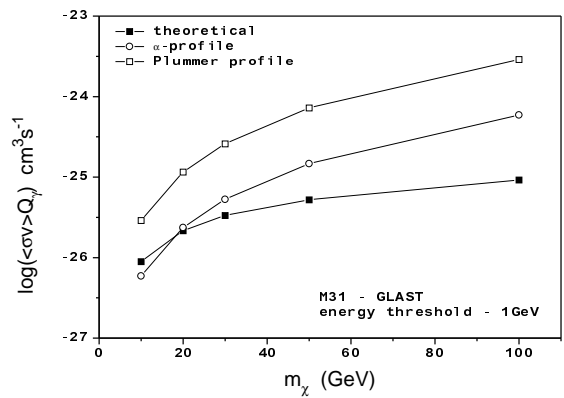}}
}
\vspace{0.2cm}
\caption
{Minimum photon production 
rate required for detection and theoretical estimates 
as a function of the neutralino mass
and for an energy threshold of $1.0$ GeV. Minimum production rates
are shown for the halo of M31 with cusp ($\alpha$-profile) and
core  (Plummer) profiles.}
\label{m31glast2}
\end{figure}

\subsection{Fluxes from Dwarf  Spheroidals}

Accumulating radial velocity data on stars of dwarf spheroidal (dSph)
in the past years, have been providing convincing support for a large M/L ratios and hence 
a substantial amount of dark matter in these galaxies. The possibility of 
detecting  $\gamma$-ray emission
from dSph's due to neutralino annihilation has been considered in 
some recent works \cite{bbsts1999,efs2003}, most of which 
emphasize on the energy range covered by ACTs.
In the photon energy range of around $0.1$ GeV, the acceptance
angle of GLAST covers a considerable fraction of
the galaxy, hence increasing the amplitude of the received signal.

Once again, the predicted flux depends on the dark
matter density profile
of dSph's. The light distribution in these systems is well-described by 
isotropic King models 
\cite{king1962}. However, in the case of Draco the velocity dispersion rises slightly with
the increasing radius, ruling out the possibility that the DM distribution be similar
to that of light \cite{kweg2001}. The observed velocity dispersion profile can be
reproduced by models in which the dark matter distribution is significantly more extended
than light, having also a tangential anisotropy in the velocity dispersion
 \cite{kweg2001}.
These models are based on a modified Plummer potential and density family, characterized 
essentially by  three parameters: $r_0, \beta$, and $\gamma$. The first 
defines the core or the
scale radius, the second determines the mass 
distribution ($\beta$ = 1 implies that DM follows
light) and the third measures the velocity anisotropy.

The potential is given by \cite{wkeg2002}
\begin{equation}
\phi(r) = \phi_0 (1 + \frac{r^2}{r_0^2})^{\beta/2}
\label{plummer2}
\end{equation} 
where $\phi_0$ is related to the central  matter
density by the equation $\rho_0 = 3\mid\beta\mid\phi_0/(4\pi G r_0^2)$. The central
density can  also be explicited in terms of the projected central velocity 
dispersion $\sigma$ as \cite{wkeg2002}
\begin{equation}
\rho_0 = \frac{\sigma^2}{\pi^{3/2}Gr_0^2}F(\beta, \gamma)
\label{rozero}
\end{equation}
where $F(\beta, \gamma) = (5 + \beta - \gamma)\Gamma(5/2+\beta/2)/\Gamma(2+\beta/2)$. 

According to \cite{kweg2001}, the best parameters describing the Draco system are:
$r_0$ = 9'.6 (0.2 kpc at a distance of 72 kpc), $\beta$ = -0.34 and $\gamma$ = -1.03
(implying tangential velocity anisotropy). Since the observed central projected velocity
dispersion is 8.5 km.s$^{-1}$, from (\ref{rozero}) one obtains for the central density
$\rho_0 = 3.8\times 10^{-23}$ g.cm$^{-3}$.

We have used this modified Plummer model and the parameters above to estimate
the $\gamma$-ray flux from Draco. If we adopt $m_{\chi}$ = 30 GeV, corresponding
to the lower limit derived from accelerator data, the expected $\gamma$-ray flux
from Draco above 0.1 GeV is about 1.3$\times 10^{-10}$ cm$^{-2}$.s$^{-1}$, which
is about one order of magnitude less than the sensibility of GLAST at this energy
threshold.  Higher neutralino masses  will give still lower fluxes. A similar result is 
obtained for 1.0 GeV energy threshold.

The situation is more complicated in the
case of Sagittarius since this dSph is partially disrupted by the tidal forces.
Up to now, the different proposed models are not able to reproduce adequately
the age and the observed structure of this dSph. Most of the models suggest
that the system is disrupted after 1-2 orbits while observations indicate ten or more.
  
Numerical simulations aiming to describe the merger history of this dwarf
assumed either a Plummer  \cite{majewskietal2003} or a King profile
\cite{hw2001} for the progenitor. These studies suggest
that the mass of the core is in the range
$(5 - 20) \times 10^8\, {\rm M}_{\odot}$ \cite{ibataetal1997,majewskietal2003,hw2001}. 

Here we model the mass distribution of Sagittarius by an isotropic 
King profile, since no significant
velocity dispersion gradient is detected across the main body of the galaxy.
The tidal radius is ill-defined in this
system, but it should be smaller than $4$ kpc in order to account for the
$2-3$ Gyr old M giants
which were formed in the core and then escaped through the tidal boundary. If we take
respectively for the tidal and core radii  $r_t = 2$ kpc and $r_c = 0.55$ kpc, the total 
mass is about $1.8 \times 10^9\, {\rm M}_{\odot}$, consistent with simulations by
\cite{hw2001}. Since the central projected velocity dispersion is 11.4 km.s$^{-1}$
 \cite{ibataetal1997}, the
resulting central density is $\rho_0 = 0.03\,\,{\rm M}_{\odot} {\rm pc}^{-3}$.
In Fig.\ \ref{dwarfs} we show the minimum photon production rate required
for detection and the theoretical rate as a function of the neutralino mass.
Emission from Sagittarius
can be detected by GLAST above $0.1$ GeV, if the neutralino mass is $\leq 50$ GeV, but
not above $1$ GeV, due to the small acceptance angle at these energies.
\vspace*{-0.4cm}
\begin{figure}
\centerline
{
\epsfxsize=0.5\textwidth
{\epsfbox{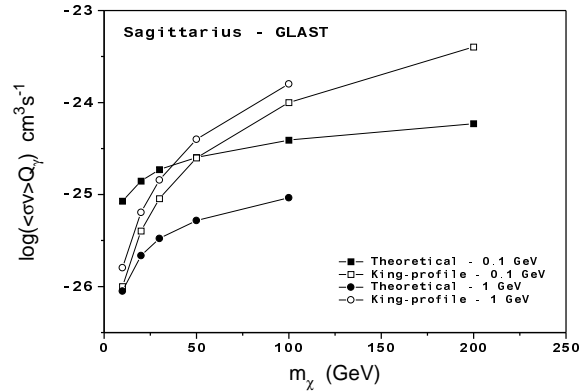}}
}
\vspace*{0.2cm}
\caption
{Minimum photon production rates and theoretical rates for energies above
$0.1$ GeV and $1.0$ as a function of the neutralino mass.}
\label{dwarfs}
\end{figure}

\section{Effects of clumps on the flux}
\label{clumps}

Numerical N-body simulations show that dark
matter halos contain a large number of self-bound substructures, which correspond to 
about $5-15 \%$ of their total mass \cite{moetal1999,klypinetal1999}. These
substructures are the consequence of the capture of
small satellites, which have not yet been disrupted by tidal
forces. Since the 
$\gamma$-ray emissivity depends on the square of the density, these substructures
enhance the $\gamma$-ray flux relative to that expected  for a smooth
halo. However, whether this enhancement is significant or not is still a
matter of debate. 
In some simulations, an important
enhancement of $\gamma$-ray flux due to clumpiness was
reported for emission from our halo
\cite{cm2000}, while other numerical simulations
seem to have indicated only a very small boost of the signal \cite{swsty2003}.
The enhancement due to clumps not only depends on the density profile of the
clumps but also on their number density.
The predicted number of massive substructures
exceed the number of observed satellites of the Milky Way or M31, by at least one order of
magnitude, which has been causing severe problems for $\Lambda$CDM model. For
instance, a number of subhalos with masses above $10^8$ M$_{\odot}$ as high as
$500$ has been assumed in recent works \cite{to2002}.

Here, to study the effects of clumpiness and the effects of the density
profile of clumps on the flux, we have made many different experiments
with many {\it test halos} that we have produced numerically. In this way,
we have been able to study the effects of clumpiness for halos with very high
resolution and with various different profiles.
Our test halos have the characteristic parameters of M31, with
typical mass of $M_h =1.5 \times 10^{12}$ M$_{\odot}$, 
constituted by $5\times 10^7$ particles, corresponding to a mass resolution of
$3 \times 10^4$ M$_{\odot}$. 
In the first experiment, the particles are distributed in a spherical volume of radius $250$ kpc,
comparable to the gravitational radius expected for halos of such a mass (see \cite{pmp2003}),
according to $\alpha$-profile. This will be our {\it reference halo}.
In the next trials, we consider halos with substructures. The number of clumps 
N$_{cl}$ in the mass range $m$ - $m$+$dm$ is assumed to obey \cite{moetal1999,swsty2003}
\begin{equation}
dN_{cl} = \frac{A}{m^{1.78}}dm\,.
\label{clumpsequation}
\end{equation}
The normalization constant $A$ is calculated by requiring the total mass in the clumps
to be $10\%$ of the halo mass and by assuming subhalos masses in the range
$10^{5.5}$ -- $10^{9.25}$ M$_{\odot}$. The minimum number of particles in the clumps is
about $14$, while the more massive ones have $45,000$ particles. These figures
correspond to mass bins built as described below. In spite of the low number
of particles in the small clumps, our results are not seriously affected, since most
of the flux enhancement is due to the massive structures. For numerical convenience, the
number of clumps was calculated from (\ref{clumpsequation}) within logarithmic bins of
width equal to $0.25$. 

Clumps were distributed according to the (normalized)
probability distribution
$p(r)d^3r = (\rho( r)/M_h)d^ 3r$, where $\rho( r)$ is assumed to
have an $\alpha$-profile, which gives the probability to
find a clump at a distance $r$ within the volume element $d^3r$. 
The stripping process caused by tidal forces
seems to reduce the density of a clump at all radii and, in particular, in the central
regions, producing a density profile with a central core \cite{hayetal2003}.
This result was further confirmed by simulations which found that the inner structure of
subhalos are better described by density profiles shallower than NFW \cite{swsty2003}.
However, other simulations seem to indicate that the central regions of 
clumps are well-represented by 
power law density profiles, which remain unmodified 
even after important tidal stripping \cite{kmmdsm2003}. 
To test the importance of these differences for the flux, we
allow for both of these possibilities: we simulate both test halos with clumps having a
central core
$\rho(r) \propto 1/(r_0 + r)^2$ or NFW profile and study how 
our results would change from one to the other.

\begin{figure}
  \begin{center}
    \rotatebox{0}{ \includegraphics[height=6cm,width=8cm]{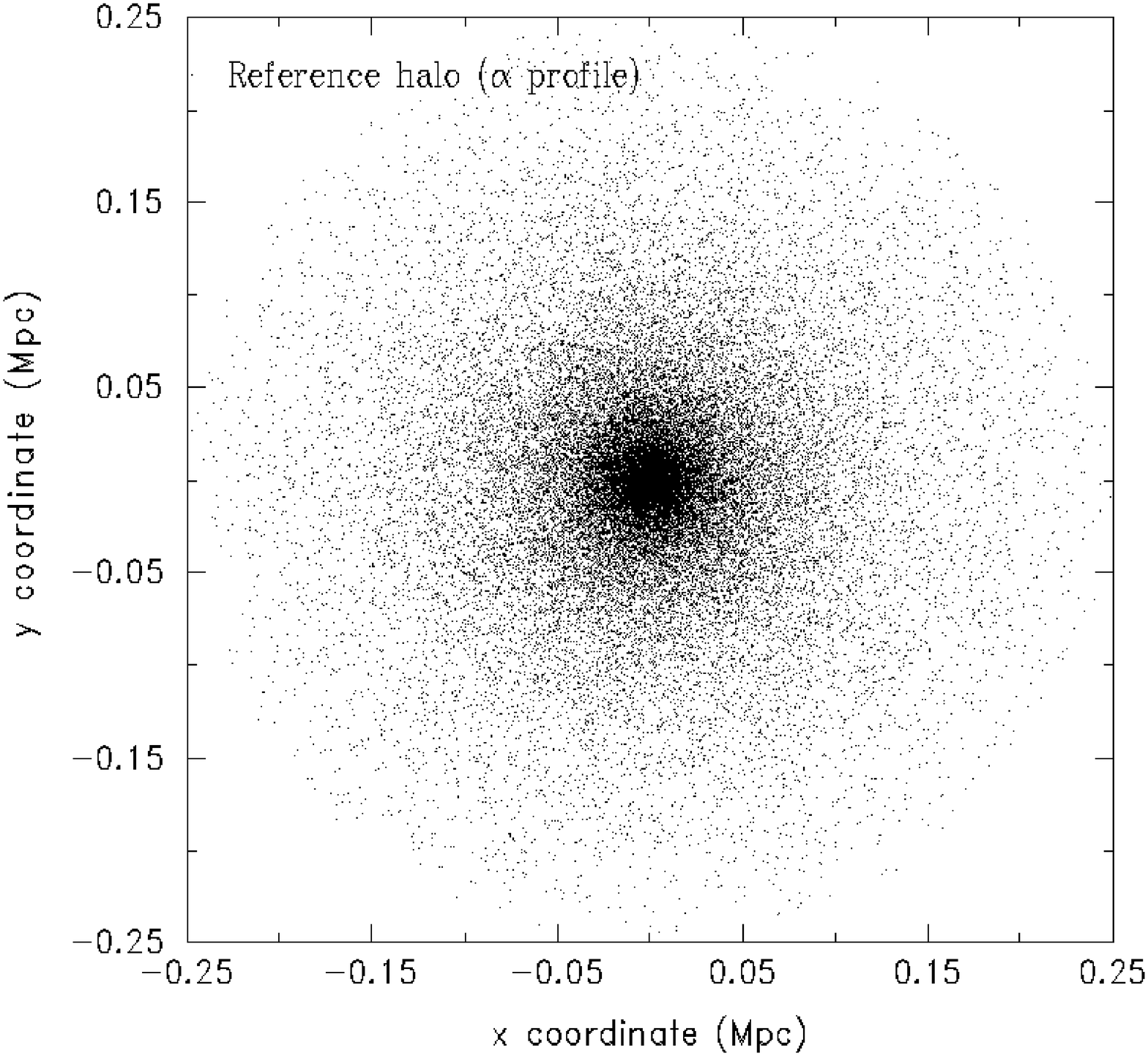}}
  \end{center}
   \begin{center}
    \rotatebox{0}{ \includegraphics[height=6cm,width=8cm]{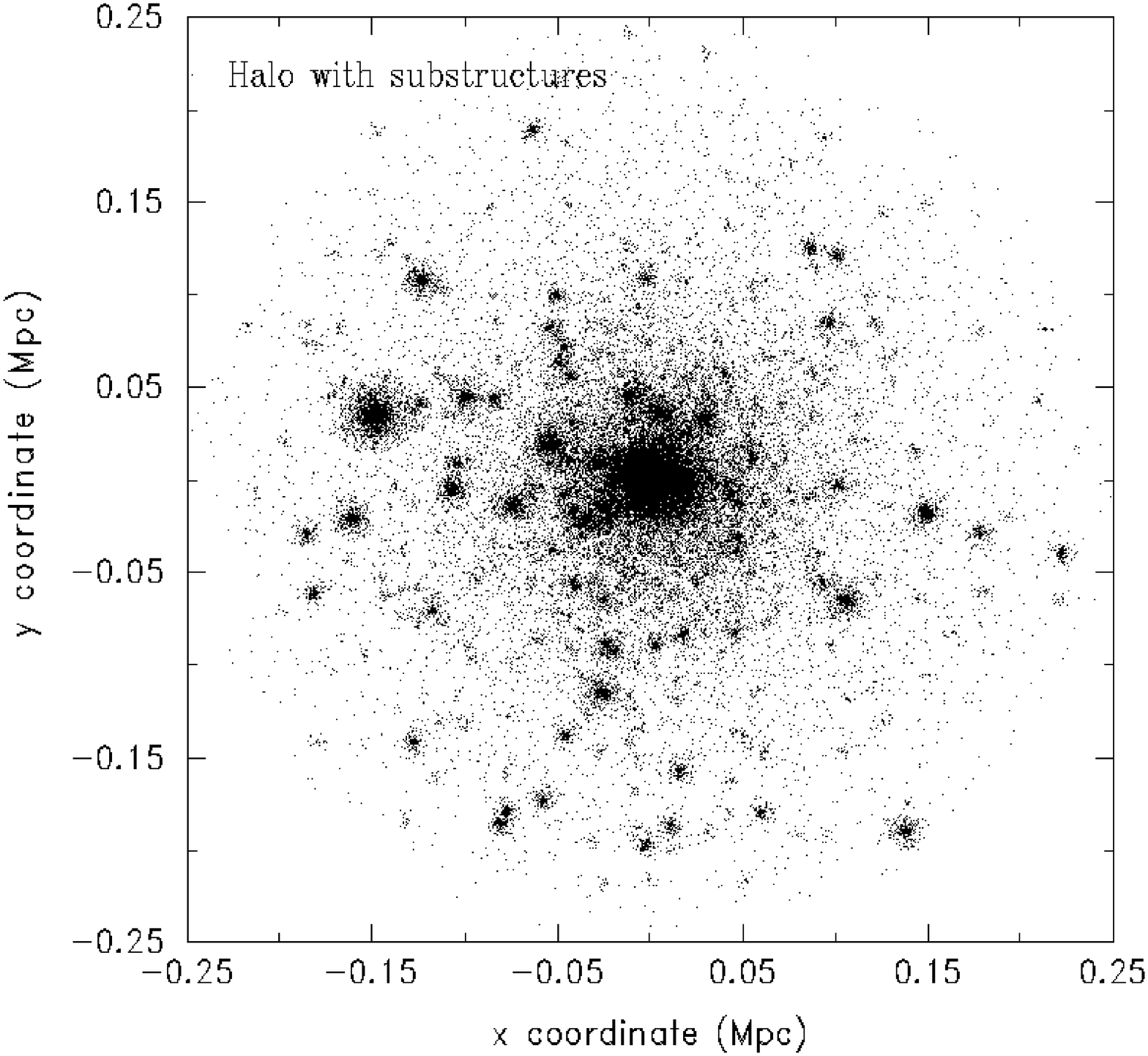}}
  \end{center}
\caption
{Two-dimensional thin slices of the halos containing 5$\times 10^7$ particles, which
we have numerically modeled to study effects of clumpiness on the $\gamma$-ray
emission. The top panel shows one of our reference halos, which has the
characteristics of M31, with an $\alpha$-profile and
no substructures.
The bottom slice shows a halo with  $10\%$ of its mass distributed 
in substructures which, in the particular 2-dimensional plot show here,
have NFW profiles. The number of substructures is 
taken from the power-law (\ref{clumpsequation}) and
their centres are distributed according to an $\alpha$-profile. The clumps are 
highlighted by reducing  the number of points of the background halo particles.
}
\label{subhalos}
\end{figure}

We recall that the $\gamma$-ray intensity (\ref{igamma}) depends on the 
the integral along the line of sight of the square of the density.
We define the enhancement factor to be the ratio between the
reduced intensity calculated for halos with substructures to that calculated
for our reference halo (with no substructure).. The numerical evaluation of intensities was performed by
dividing the halo volume in a large number of small cubes. The cube volume
varies logarithmically, being smaller in the central regions in order to increase
the accuracy in the particle density. For angular distances $\theta > 1^o$, the
number of cubes is about $10^6$ whereas near the centre ($\theta < 1^o$)
this number increases up to $15 \times 10^6$.
We have checked our algorithm for computing the reduced intensity 
${\cal I}$ by comparing from direct integration
of (\ref{alpha}). Errors are of the order of 2-3 \%  if the integration  is
carried along line of sights far from the centre, but can reach values of up to
10\% if the integration is performed along directions close to the centre,
due to resolution problems. Intensities were calculated
in steps of $\Delta\theta = 0.1^o$ and $\Delta\psi = 1.0^o$, with $\theta$ and
$\psi$ being respectively the angular distance to the centre and the azimuthal
angle around the centre of the galaxy.

\begin{figure}
  \begin{center}
    \rotatebox{0}{ \includegraphics[height=6cm,width=8cm]{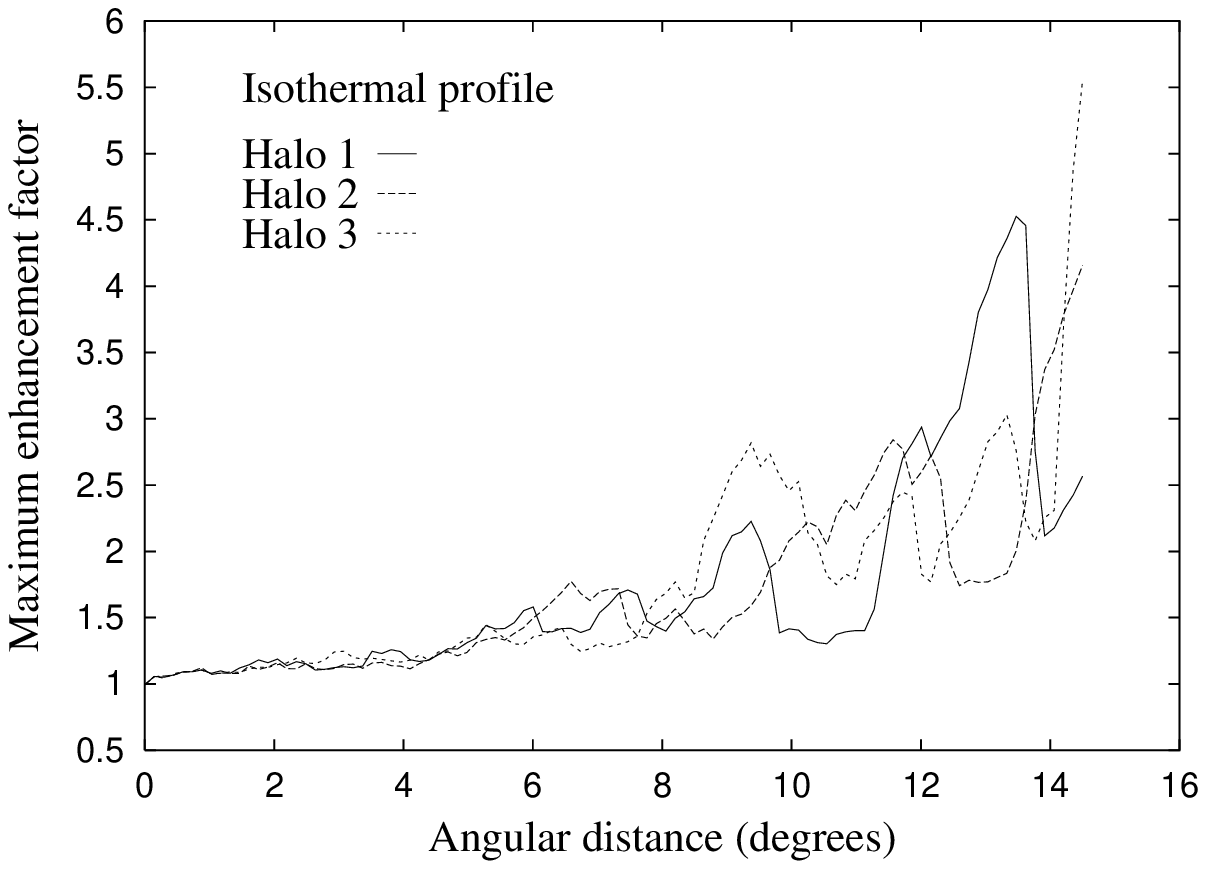}}
  \end{center}
   \begin{center}
    \rotatebox{0}{ \includegraphics[height=6cm,width=8cm]{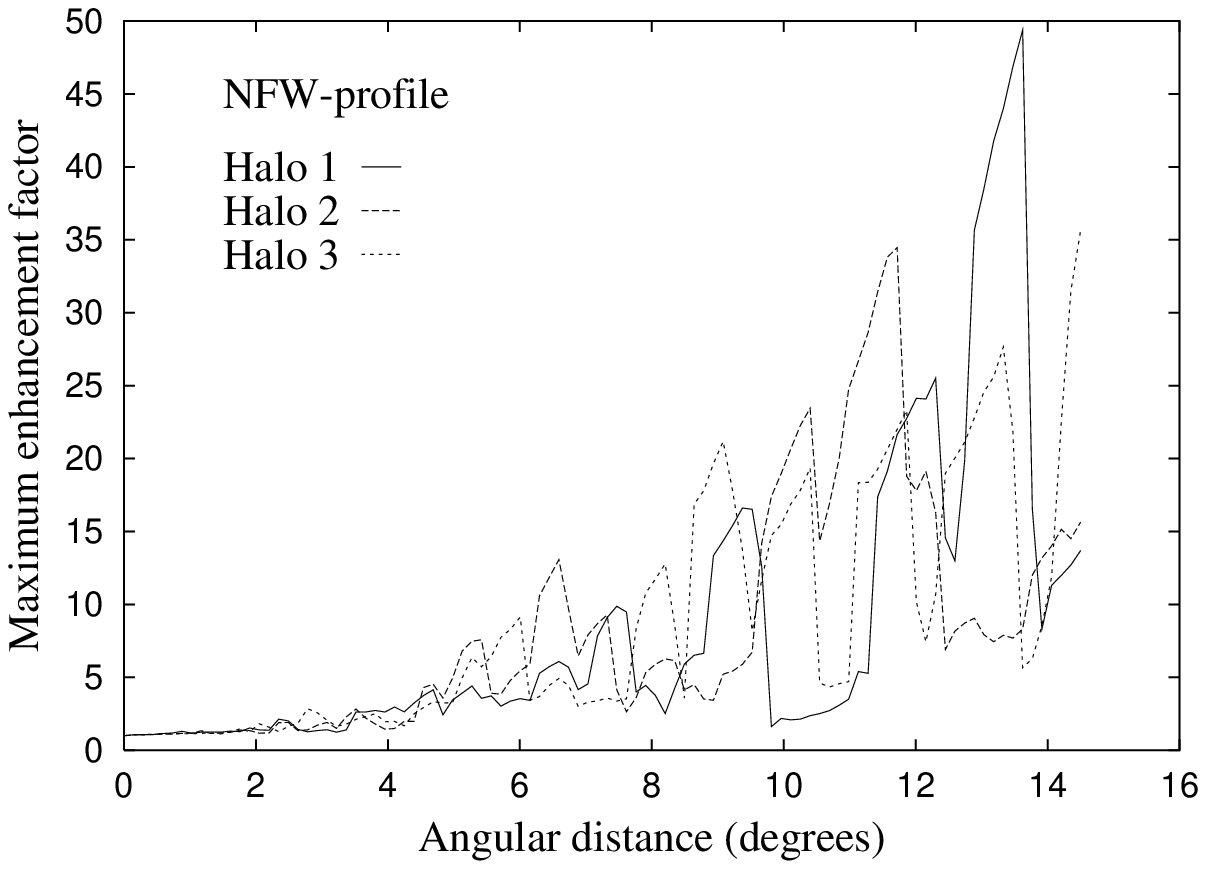}}
  \end{center}
\caption
{Plots show the variation of the enhancement factor  as
a function of the angular distance (in degrees) from the centre, for halos
with characteristics of M31. Clumps have either an isothermal (top plot) or
a NFW profile (bottom plot). In each
case, three different simulations are shown to illustrate  
statistical fluctuations.}
\label{subhalos}
\end{figure}

In figure \ref{subhalos} we show the {\it maximum} enhancement factor as a function 
of the angular distance to the centre of M31 for some selected trials. In the upper
panel clumps have profiles with central cores whereas in the lower panel a profile with
central cusp
was adopted. We notice that, up to distances of about $4^o-5^o$, higher than
the resolution of GLAST, independent of the clump profile, the enhancement is
very small (about 10\%), in agreement with the previous 
results ({\it e.g.} see \cite{swsty2003}). Larger enhancements
may be obtained at the outskirts of the halo and, in this case, clumps with
central cusps
produce a strengthening almost one order of magnitude higher than clumps with
cores. However,
two points must be emphasized: a) these enhancements are local and when convolved
with the PSF of the detector, they will be attenuated; b) as shown in 
Fig.\ \ref{intensityprofile}, at $1^o$ from the centre the intensity 
has already decreased by almost four orders of magnitude. Thus, at the halo
boundaries, even with an enhancement factor of $40-50$, the expected flux
received by a detector pointing away from the centre will be below the
detectability  limit.
\section{Effects of central supermassive black holes}
\label{blackhole} 

Fast-growing observational evidences indicate that most elliptical galaxies and bulges
of spirals harbor  SMBHs at their centres. In particular, recent spectroscopic data
suggest the existence of SMBH in M31, M87 and in the Milky Way \cite{koge2001}. These
kinematical data, if interpreted in terms of Keplerian motions, indicate a mass of
$4.5 \times 10^7$ M$_{\odot}$ for the SMBH in the centre of M31 \, \cite{koge2001}
and of about $2.8 \times 10^9$ M$_{\odot}$ for that in the centre of M87 
\cite{harmsetal1994,maccheto1997}. Kinematical data on stars very near Sagittarius A,
the radio source supposed to be in the galactic centre, represent the best evidence in favor
of the existence of a SMBH in the Milky Way \cite{ghezetal2000}, whose mass is 
$2.6 \times 10^6$ M$_{\odot}$.

If a black hole grows inside a dark  matter halo, supposed to be constituted by a
dissipationless fluid, the density around it also grows since the gravitational
attraction of the black hole causes a shrinking of the orbits inside the sphere of
influence, whose radius is given by $ r_{\rm bh} = GM_{\rm bh}/\sigma^2$, where
$\sigma$ is the initial one-dimensional velocity dispersion of dark matter. If the growth is adiabatic
\cite{young1980}, that is if $t_{\rm P} << M_{\rm bh}/\dot M_{\rm bh} <<
t_{\rm R}$, where
$t_{\rm P}$ is the orbital timescale, $\dot M_{\rm bh}$ the black hole accretion rate and
$t_{\rm R}$ the relaxation timescale, then a central spike will be formed inside the sphere
of influence, having a central cusp profile of the form 
\begin{equation}
\rho_{\rm sp} = \frac{A}{r^{\gamma}}
\label{spike}
\end{equation}
The value of the exponent $\gamma$ depends on the initial state of the matter
distribution. On the one hand,
if the initial density profile has a central core,  then the resulting spike is a power law
with $\gamma$ = 3/2 \cite{peebles1972,young1980}. On the other hand, an initial
power law profile with an exponent $\gamma_0$ will produce a steeper density profile with
an exponent $\gamma = 2 + 1/(4-\gamma_0)$ \cite{gs1999}. 

The adiabatic scenario is questionable: firstly because the predicted
accretion rates are considerably higher than the current estimates \cite{kzs2001}
and secondly, because massive galaxies like M31, M87 and the Milky Way
have most probably experienced several merging episodes \cite{pmp2003}, which
may have perturbed or even destroyed the central spike. Nevertheless, we will assume
in our calculations that there is a putative spike formed around the SMBH under
adiabatic conditions.
 
For M31, previous simulations \cite{falvardetal2002} gives a spike profile $\rho_{\rm sp} \propto
r^{-1.5}$ which is formed in less than $10^6$ yrs. According to \cite{falvardetal2002} this 
profile gives a small
enhancement of about 45\%, with respect to a NFW profile, of the $\gamma$-ray 
emission from M31.
Here we adopt the following procedure. We assume a smooth transition at 
$r_{\rm bh}$ from the spike to the background profile $\rho_{\rm h}$, which is given either
by the $\alpha$-profile  or by the Plummer profile, {\it i.e.}
\begin{equation}
\frac{A}{r_{\rm bh}^{\gamma}} = \rho_{\rm h}(r_{\rm bh})
\end{equation}
a condition which allows us to determine the constant $A$. Since the power
law profile diverges at the centre, we must  introduce a cutoff radius $r_{min}$
fixed either by the self-annihilation rate or by capture into the black hole. Using
the former condition, the cutoff radius is given by 
\begin{equation}
r_{\rm min} = (\frac{A<\sigma_{\chi\bar\chi}v>t_{\rm bh}}{m_{\chi}})^{1/\gamma}
\end{equation}
where $t_{\rm bh}$ is the age of the SMBH, here taken to be equal to $10^{10}$ yrs.
For M31 the radius of the sphere of influence is about $7.7$ pc. We have
first considered a profile with a central core given by the Plummer profile 
 for the background, which
results in a spike profile, $\rho_{\rm sp} \propto r^{-1.5}$. The $\gamma$-ray
flux was computed as before from (\ref{fgamma}), using (\ref{spike}) in  the
interval $r_{\rm min} \leq r \leq r_{\rm bh}$ and the Plummer profile (\ref{plummer}) for $r >
r_{\rm bh}$.
No significant enhancement is obtained whether the threshold energy is
taken at $0.1$ GeV or at $1.0$ GeV. However a different result
is obtained if we use the $\alpha$-profile (\ref{alpha}) as the background profile. The spike profile
is much steeper ($\gamma \approx 2.3$), causing a more rapid increase of the
density towards the centre and an important enhancement of the $\gamma$-ray flux.
The strengthening of the signal is about a factor of $38$ at $0.1$ GeV and 
increases up to a factor of $55$ at $1$ GeV, due to the higher spatial resolution of 
GLAST at higher energies. As a consequence, if the halo of M31 has a
profile with a central cusp, the $\gamma$-ray emission can be detected by GLAST at
$0.1$ GeV if $m_{\chi} \leq  300$ GeV and at $1.0$ GeV if $m_{\chi} \leq 500$ GeV.
These limits are considerably larger than those derived in Section \ref{numericalresults}
and open new possibilities to impose bounds on the neutralino mass and/or on the
dark matter distribution.

The effects of a central spike are more dramatic in M87, since this galaxy has an
extremely massive central black hole.  The radius of influence of this SMBH is about
$97$ pc. As we have seen in Section \ref{ASTRO}, the dark  matter profile of M87
can be quite well represented by the $\alpha$-profile (\ref{alpha}). Thus, the central spike has also
a profile $\rho_{sp} \propto r^{-2.3}$, but the greater dark matter mass inside
the sphere of influence produces an enhancement larger than that obtained for M31.
The boost factor is about $200$ at 0.1 GeV and $234$ at 1.0 GeV. These large
factors lead to detectable signals either at $0.1$ GeV if $m_{\chi} \leq 60$ GeV
or at $1.0$ GeV if $m_{\chi} \leq 100$ GeV.

\section{Conclusions}
\label{conclusion}

Observational results favor a Universe whose main matter
content is non-baryonic and dark. Presently, the most plausible DM candidate
is neutralino; the lightest stable supersymmetric electrically neutral Majorana fermion.
The possibility that neutralinos might be detected indirectly through 
their annihilation products such as $\gamma$-rays and neutrinos 
are explored by increasingly more sophisticated 
experiments which mainly search for energetic neutrinos from the centre of the
Earth or the Sun and or for $\gamma$-rays from the galactic halo and various extragalactic
sources. Although, the goal of DM detection has not yet been achieved, upper bounds have
been set on parameters such as neutralino mass.
  
In this work, the  $\gamma$-ray emission from neutralino annihilation
has been studied. Our aim has been to use the results of indirect search experiments
mainly GLAST to constrain the neutralino mass and the halo density
profiles as best as possible.
Hence, we have used a simple description of the annihilation process in which
the only free parameter is the neutralino mass. We have estimated the
photon production rate by fixing the present relic neutralino density
to its equilibrium value at the decoupling temperature ($\Omega_{\chi}$ =
0.26) favored by the latest CMB data and
by assuming a QCD jet description of the annihilation process. Photon production rates
above $0.1$ and $1.0$ GeV are in the range $10^{-26}$ up to $10^{-24}$ ph cm$^3$s$^{-1}$
for the considered range of masses ($10 \leq m_{\chi} \leq 2000$ GeV),
consistent with results derived from supersymmetric codes such as DarkSUSY and
SUSPECT. The mass interval that is explored here is
compatible with the lower bound ($\sim$ 30 GeV) given by the accelerator data and with
the recent upper limit ($\sim$ 800 GeV) set by the elastic scattering of neutralinos from
protons, including constraints imposed by new measurements of the anomalous magnetic
moment of the muon ($g_{\mu}$-2) \cite{knra2002}. 

The DM density profile, was studied in the light of
the different spatial resolution of the future experiment GLAST at different energies.
Two different dark matter density profiles were adopted in our calculations: a
profile with central cusp,
in agreement with results of numerical simulations and one with a central core, more representative
of the observational data on rotation curves of bright spirals. These profiles are characterized
by two parameters determined by the halo mass and by the mass density of dark matter at a given
distance to the centre of the halo (in the case of the the Milky Way and M31) or using data on the
X-ray emissivity (in the case of M87). For dSphs, a modified Plummer profile, able to account for
the observed raising velocity dispersion profile, was adopted for Draco whereas
a King profile was assumed for Sagittarius.

We have studied, by generating and experimenting with many different
numerically simulated halos, the effect of clumpiness on the $\gamma$-ray 
emission from M31 and conclude that the enhancement is less than 
a factor of two for
detectors centered on the object and having acceptance solid angles of $10^{-5}- 10^{-3}$
sr$^{-1}$. The local enhancement factor can  be as large as 40-50 at the boundaries
of halos with central cusps.  However, the absolute fluxes expected
for a detector pointing away from the centre will be, in general, below the limit of
detectability. These conclusions are in agreement with  studies on the clumpiness
effects either in the galactic halo \cite{swsty2003,kzw2004} or in M31 \cite{pb2003},
where maximum enhancement factors of about a factor of three were obtained.

Our calculations show that if the dark matter halo of the Milky Way has a density
profile with a central core, then
the expect $\gamma$-ray intensity above 1 GeV in the direction $b = 90^o$ 
and averaged over a solid angle of  $0.5^o\times 0.5^o$, is comparable to the EGRET bounds
\cite{dixonetal1998}, if $m_{\chi} \leq$ 50 GeV. However, a profile with a central cusp
gives an expected intensity at least one order of magnitude smaller, requiring an
important enhancement due to the halo clumpiness to explain the data. This
could be an additional argument in favor of high baryon-to-dark matter ratios 
in the central regions
of bright galaxies, in agreement with  analyses of rotation curves.

Different spatial resolution at different energies of GLAST permits us to obtain
informations either on the neutralino mass or on the dark matter
distribution in the halos. Combining the informations on both energy 
thresholds, four possibilities arise which are written below.

For M31, including effects of the spike produced by the central SMBH,
the following situations may occur: 

\noindent
a) If no detection is made at either energies then one may draw the conclusions
that $m_{\chi} \geq 20$ GeV {\it and} that the profile has a central core.

\noindent
b) If detection is made at both energy thresholds then the profile has a
central cusp {\it and} $m_{\chi} \leq 300$ GeV.

\noindent
c) If detection is made at $0.1$ GeV but not at $1.0$ GeV then we may conclude
that the profile has a central core {\it and} $m_{\chi} \leq 20$ GeV.

\noindent
d) Detection at $1.0$ GeV but not at $0.1$ GeV would result from a central
cusp in the density profile {\it and} $300 \leq m_{\chi} \leq 500$ GeV.

A similar analysis for M87 yields:

\noindent
a) If no detection is made at either energies then one may draw the conclusions that 
the "$\alpha$-profile" is not valid at the centre where the density profile
probably has a core or $m_{\chi} \geq 100$ GeV.

\noindent
b) If detection is made at both energy thresholds then
the "$\alpha$-profile" is a good representation of the dark matter distribution near
the centre and $m_{\chi} \leq 60$ GeV.

\noindent
c)  If detection is made at $0.1$ GeV but not at $1.0$ GeV then we may conclude
that profile has a central core but no limits can be set on the neutralino mass
unless an specific density profile is adopted.

\noindent
d)  Detection at $1.0$ GeV but not at $0.1$ GeV would indicate that
the "$\alpha$-profile" is acceptable and $60 \leq m_{\chi} \leq 100$ GeV.

Because of their high dark matter content 
and their relative proximity the dSph Draco and
Sagittarius are among the most favorable sources of 
$\gamma$-rays resulting from neutralino annihilation. 
However, we conclude that in spite of its large M/L ratio, the expected signals from Draco
will not be detectable in the energy range of GLAST. On the contrary,
our results suggest
that an emission detectable by GLAST
above $0.1$ GeV is expected for Sagittarius, if the neutralino mass is below $50$ GeV, a
limit compatible with the LEP lower bound.

S.\ P.\ acknowledges PhD fellowship from Universit\'e de Nice Sophia-Antipolis
(UNSA). R.\ M.\ is supported by Marie
curie fellowship HPMF-CT 2002-01532.
We thank F. Stoehr, S. Profumo and the referee for useful comments which
have improved the manuscript.

\end{document}